\documentclass[amsmath,amssymb,floatfix,preprint,showkeys]{revtex4}

\usepackage{graphicx}
\usepackage{dcolumn}
\usepackage{bm}

\begin{document}

\title{Applying the extended molecule approach to correlated electron transport: important insight from model calculations}

\author{Ioan B\^aldea}
\email{ioan@pci.uni-heidelberg.de}
\altaffiliation[Also at ]{NILPRP, ISS, RO 077125 Bucharest, Romania.}
\author{Horst K\"oppel} 
\affiliation{Theoretische Chemie,
Physikalisch-Chemisches Institut, Universit\"{a}t Heidelberg, Im
Neuenheimer Feld 229, D-69120 Heidelberg, Germany}

\author{Robert Maul}
\author{Wolfgang Wenzel}
\affiliation{Institut f\"ur Nanotechnologie, Forschungszentrum Karlsruhe, 
D-76021 Karlsruhe, Germany}

\begin{abstract}
Theoretical approaches of electric transport in correlated molecules 
usually consider an extended molecule, which includes, in addition to the 
molecule itself, parts of electrodes. In the case where electron correlations 
remain confined within the molecule, and the extended molecule is sufficiently large, 
the current can be expressed by means 
of Laudauer-type formulae. Electron correlations are embodied into 
the retarded Green's function of a sufficiently large but isolated extended molecule,
which represents the key quantity that can be accurately determined by means of ab initio
quantum chemical calculations. To exemplify these ideas, we present and analyze 
numerical results obtained 
within full CI calculations for an extended molecule described by the 
interacting resonant level model. Based on them, we argue that for narrower band (organic) electrodes 
the transport properties can be reliably computed, because 
the extended molecule can be chosen sufficiently small to be tackled 
within accurate ab initio methods. For wider band (metallic) electrodes, larger extended molecules 
have to be considered in general, but a (semi-)quantitative description of the transport 
should still be possible in the typical cases where electron transport 
proceeds by off-resonant tunneling.
Our numerical results also demonstrate that, contrary to the usual claim,
the ratio between the characteristic Coulomb strength and the level width due to molecule-electrode 
coupling is not the only quantity needed to assess whether electron correlation effects are strong or weak.
\end{abstract}

\keywords{molecular electronics, correlated molecular transport, molecular orbital gating, 
electrode-molecule contacts, nonequilibrium Green's functions, Keldysh formalism, interacting resonant level model}
\maketitle
\section{Introduction}
\label{sec-introduction}
Intensive experimental and theoretical work done in the last decade led to significant advances 
in the field of electric transport through molecules and nanostructures
\cite{Goddard:07,Kim:09}. From the theoretical side, ab initio approaches --- mostly based on
ground state density-functional theory (DFT) calculations
combined with the Keldysh nonequilibrium Green's functions (NEGF) --- 
contributed to a qualitative understanding of some trends important for molecular 
electronics \cite{Wenzel:03b,Stokbro:03}. 
However, intrinsic limitations of the DFT-approaches make it 
impossible to quantitatively 
describe electronic transport in  correlated molecular systems. 
There is a broad consensus in attributing
the rather large quantitative discrepancies between the calculated and measured currents
to the notorious fact that the DFT substantially underestimates the excitation energies, 
in particular the HOMO-LUMO gap. In order to more appropriately account for electron 
correlation effects on nanotransport, several studies, which also used the NEGF, 
focused their attention on other methods, e.~g., the so-called GW approximation 
\cite{Darancet:07,Thygesen:07,Thygesen:08b,Thygesen:08c,Millis:08,Millis:09}, 
a method proposed long time ago for metals \cite{Hedin:65}.

The theory of electric transport through correlated molecular systems (i.~e., 
which cannot be described within a single-particle picture) 
is confronted with problems, which are not encountered in the absence of 
electron correlations. A basic issue is that of partitioning the real 
system into electrodes and the molecule \cite{Jauho:06}. 
In the uncorrelated case, the electric current can be computed \emph{exactly} 
within the Landauer approach. Being exact, the result 
is partition-independent, and therefore the specification of the electrode-molecule 
interfaces used in calculations is not important. For physical reasons, the current 
through a correlated molecular 
system should also be partition-independent. Excepting for a few models, 
the calculations on the transport in correlated systems are not exact. The
various approximation schemes employed do not necessarily satisfy this partition invariance.

We are not aware of a systematic study addressing this issue. 
In most cases, the reported results are from calculations carried out for the 
largest extended molecule that can be tackled, wherein besides the molecule itself 
the largest possible electrode portions linked to it are included.
The main concern of the most elaborate 
theoretical approaches of the correlated molecular transport developed so far 
is not the partition invariance or, alternatively, the independence 
on the size $N$ of the extended molecule.
One is merely interested whether the employed approximate approach is current 
conserving or not, 
i.~e., whether the calculated steady-state current does depend or not 
on the position in the extended molecule of the cross-sectional area through which it flows 
\cite{KadanoffBaym,Danielewicz:84,Thygesen:08b,Petri:09}.

\begin{figure}[htb]
\centerline{
\includegraphics[width=0.25\textwidth,angle=-90]{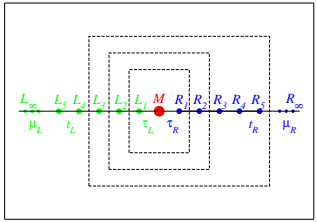}}
$ $\\[2ex]
\caption{\label{fig:setup} 
Schematic representation of a typical two-terminal 
setup for molecular transport. A molecule $M$ is linked to two left $L$ and right $R$ 
electrodes. In a correct approach, the steady-state current is site-independent and 
does not depend on the way in which the total system is partitioned into the central 
part (extended molecule) and the electrodes. In the figure, the different 
partitions are depicted by the dashed-line frames.}
\end{figure}

To summarize, a reliable approach of the transport in correlated molecular systems 
should not only yield a steady-state current 
that does not depend on the transverse areas at $\ldots L_3, L_2, L_1, M, R_1, R_2, R_3, \ldots$,
but also on the way in which the extended molecule is delimited by the various rectangles, 
as schematically depicted in Fig.~\ref{fig:setup}.\\[1ex] 

The remaining part of the paper is organized in the following manner. In Sect.~\ref{sec-theory},
we present the theoretical framework, and in Sect.~\ref{sec-large-ext-mol} 
the rationale for the concept of a sufficiently large extended 
molecule. The interacting resonant level model, which is used in all numerical calculations of this paper, 
is described in Sect.~\ref{sec-model}. Next, in Sect.~\ref{sec-conductance} we report 
numerical results for the linear conductance. 
The case of narrower band electrodes is examined in Sect.~\ref{sec-organic}, 
Sect.~\ref{sec-metal} is devoted to the case of wider band electrodes.
To be specific, for these two cases we shall use the terms organic and metallic
electrodes, respectively.
In Sect.~\ref{sec-renorm}, we present a simple renormalization scheme, which 
works surprisingly well and yields results that exhibit a very weak size dependence even 
for unrealistically strong Coulomb contact interactions. In Sect.~\ref{sec-discussion}, 
we discuss the relevance of numerical results for realistic systems, and 
in Sect.~\ref{sec-conclusion} we present the conclusions.
\section{Theoretical framework}
\label{sec-theory}
Within the Keldysh nonequilibrium Green's function framework, it is possible to express 
the steady-state current $I$ through a correlated molecule linked to electrodes, 
wherein electrons are 
supposed to be uncorrelated, by the formula \cite{Meir:92,Meir:94,HaugJauho} 
\begin{equation}
\label{eq-I-2}
I = \frac{i e}{2h}\int d\,\varepsilon\, \boldsymbol{Tr}
\left[ 
\left(f_L \boldsymbol{\Gamma_L} - f_R \boldsymbol{\Gamma_R} \right) \left(\boldsymbol{G^r} - \boldsymbol{G^a} \right)
+ 
\left(\boldsymbol{\Gamma_L} - \boldsymbol{\Gamma_R} \right) \boldsymbol{G^{<}}
\right] \ .
\end{equation}
Above, the obvious $\varepsilon$-dependence of the integrand has been omitted for brevity.
In Eq.~(\ref{eq-I-2}), $\boldsymbol{G^r}$, $\boldsymbol{G^a}$, and $\boldsymbol{G^{<}}$ 
stand for the nonequilibrium ($V\neq 0$)
exact retarded, advanced, and the lesser Green's functions of the extended molecule, 
respectively, and  $\boldsymbol{\Gamma^{L}}$
and $\boldsymbol{\Gamma^{R}}$ describe the coupling of the extended molecule 
to the left (L) and right (R) electrodes, respectively. 
The bold symbols are used to denote matrices, whose indices label the basis states of the 
extended molecule. These are site indices for the case considered below; in a realistic 
ab initio calculations, they would specify the molecular orbitals characterizing the extended molecule.
As usual, each electrode is assumed to be in equilibrium; $f_{L,R}$ represent 
Fermi distribution functions characterized by chemical potentials $\mu_{L,R}$,
which are imbalanced ($\mu_L - \mu_R = eV$) due to an applied voltage $V$ responsible for the 
current flow. 

The Green's functions $\boldsymbol{G^r}$, $\boldsymbol{G^a}$, and $\boldsymbol{G^{<}}$ 
entering Eq.~(\ref{eq-I-2}) are those of the extended molecule \emph{coupled} to the electrodes. 
To compute these nonequilibrium Green's functions one 
usually resorts to the Dyson equations \cite{Meir:94,HaugJauho,mahan} 
\begin{eqnarray}
\displaystyle
\label{eq-G-r,a}
\boldsymbol{G_{}^{r,a}} & = & \boldsymbol{G_{0}^{r,a}} + 
\boldsymbol{G_{0}^{r,a}} \boldsymbol{\Sigma_{}^{r,a}} \boldsymbol{G_{}^{r,a}} \ , \\
\label{eq-G-lesser}
\boldsymbol{G}^{\stackrel{<}{>}} & = & 
\left(\boldsymbol{1} + \boldsymbol{G_{}^{r}} \boldsymbol{\Sigma_{}^{r}} \right)
\boldsymbol{G_{0}^{\stackrel{<}{>}}} 
\left(\boldsymbol{1} + \boldsymbol{\Sigma_{}^{a}} \boldsymbol{G_{}^{a}} \right)
+ \boldsymbol{G_{}^{r}} \boldsymbol{\Sigma_{}^{ \stackrel{<}{>}  }} \boldsymbol{G_{}^{a}} 
\rightarrow 
\boldsymbol{G_{}^{r}} \boldsymbol{\Sigma_{}^{ \stackrel{<}{>}}} \boldsymbol{G_{}^{a}} \ ,
\end{eqnarray}
which relate the exact Green's functions (without subscript) to those of 
a noninteracting (uncorrelated) reference (subscript $0$) by means of the self-energies 
$\boldsymbol{\Sigma}$'s.
Uncorrelated electrons can refer either to free electrons or described within the SCF 
approximation.
The expression at the right of the arrow in Eq.~(\ref{eq-G-lesser}) applies to the 
case of a steady-state flow considered here \cite{Jauho:06}.
The above nonequilibrium Dyson equations 
can be written in a more compact form by means of the block matrices \cite{mahan}
\begin{equation}
\displaystyle
\label{eq-G-Sigma}
\boldsymbol{\hat{G}} = \left(
\begin{array}{cc} 
\boldsymbol{G^{t}} & - \boldsymbol{G^{<}}\\ 
\boldsymbol{G^{>}} & - \boldsymbol{G^{\tilde{t}}} 
\end{array}
\right) \ ; \ \ \ \ 
\boldsymbol{\hat{{\Sigma}}} = \left(
\begin{array}{cc}
\boldsymbol{{\Sigma}^{t}} & - \boldsymbol{{\Sigma}^{<}}\\ 
\boldsymbol{{\Sigma}^{>}} & - \boldsymbol{{\Sigma}^{\tilde{t}}} 
\end{array}
\right) \ ,
\end{equation}
where the causal ($t$), anti-causal ($\tilde{t}$), and the greater ($>$) matrix elements  
are related to the earlier specified ones by
\begin{eqnarray}
\displaystyle
\label{eq-G-r}
\boldsymbol{G_{}^{r}} & = & \boldsymbol{G_{}^{t}} - \boldsymbol{G_{}^{<}} = 
\boldsymbol{G_{}^{>}} - \boldsymbol{G_{}^{\tilde{t}}} \ , \\
\label{eq-G-a}
\boldsymbol{G_{}^{a}} & = & \boldsymbol{G_{}^{t}} - \boldsymbol{G_{}^{>}} = 
\boldsymbol{G_{}^{<}} - \boldsymbol{G_{}^{\tilde{t}}} = \left({\boldsymbol{G_{}^{r}}}\right)^{\dagger} \ .
\end{eqnarray}
Dyson equations equivalent to Eqs.~(\ref{eq-G-r,a}) and (\ref{eq-G-lesser}) 
relating the relevant Green's functions and self-energies can be written 
in the following compact forms 
\begin{eqnarray}
\displaystyle
\label{eq-G-0}
\boldsymbol{\hat{G}^{-1}}_{0} & = &
\boldsymbol{\hat{G}^{-1}}_{i,0} - \boldsymbol{\hat{\Sigma}}_{0,L} - \boldsymbol{\hat{\Sigma}}_{0,R} \ , \\
\label{eq-G}
\boldsymbol{\hat{G}_{}^{-1}} & = &
\boldsymbol{\hat{G}^{-1}}_{i} - \boldsymbol{\hat{\Sigma}}_{L} - \boldsymbol{\hat{\Sigma}}_{R} \ , \\
\label{eq-G-i}
\boldsymbol{\hat{G}^{-1}}_{i} & = & \boldsymbol{\hat{G}^{-1}}_{i,0} - \boldsymbol{\hat{\Sigma}}_{i, corr} \ , \\
\label{eq-G-2}
\boldsymbol{\hat{G}_{}^{-1}} & = &
\boldsymbol{\hat{G}^{-1}}_{0} - \boldsymbol{\hat{\Sigma}}_{corr} \ .
\end{eqnarray}
Here, the subscript $i$ refers to an isolated molecule.
Making use of the above equations, the Green's function matrix $\boldsymbol{\hat{G}}$ 
of the molecule linked to electrodes needed to compute the 
current of Eq.~(\ref{eq-I-2})
can be expressed in the following two equivalent forms
\begin{eqnarray}
\displaystyle
\label{eq-dyson-corr}
\boldsymbol{\hat{G}_{}^{-1}} & = &
\boldsymbol{\hat{G}^{-1}}_{i,0} - \boldsymbol{\hat{\Sigma}}_{0,L} - \boldsymbol{\hat{\Sigma}}_{0,R} - 
\boldsymbol{\hat{\Sigma}}_{i, corr} + 
\left( \boldsymbol{\hat{\Sigma}}_{i, corr} - \boldsymbol{\hat{\Sigma}}_{corr} \right) \ , \\
\label{eq-dyson-X}
\boldsymbol{\hat{G}_{}^{-1}} & = &
\boldsymbol{\hat{G}^{-1}}_{i,0} - \boldsymbol{\hat{\Sigma}}_{i, corr} -  
\boldsymbol{\hat{\Sigma}}_{0,L} - \boldsymbol{\hat{\Sigma}}_{0,R} - 
\left( \boldsymbol{\hat{\Sigma}}_{0,L} - \boldsymbol{\hat{\Sigma}}_{L} +
\boldsymbol{\hat{\Sigma}}_{0,R} - \boldsymbol{\hat{\Sigma}}_{R} \right) \ .
\end{eqnarray}
\section{The concept of a sufficiently large extended molecule}
\label{sec-large-ext-mol}
The approach exposed below relies upon the concept of a sufficiently large 
extended molecule. We shall make the \emph{physical} assumption that, 
even if, as a result of the 
molecule-electrode couplings, the electron correlations do not strictly remain confined 
within the molecule, they will be wiped out and eventually become negligible sufficiently 
far away from the molecule-electrode interfaces. 
Rephrasing, we shall consider physical situations where, although the intramolecular 
electron correlations can be strong, they rapidly decay beyond a characteristic 
length $\mathcal{L}_{c}$ and become negligible. It is the length $\mathcal{L}_{c}$ 
that defines the size of the ``sufficiently'' large extended molecule.
Being negligible beyond the sufficiently large extended molecule,
correlations insignificantly affect the embedding self-energies 
which couple the extended molecule to 
electrodes ($x=L,R$),
$\boldsymbol{\hat{\Sigma}}_{x} - \boldsymbol{\hat{\Sigma}}_{0,x} \ll \boldsymbol{\hat{\Sigma}}_{0,x}$.
That is, the expression in the parenthesis of Eq.~(\ref{eq-dyson-X}) can be ignored 
if the extended molecule is sufficiently large.
Importantly, this assumption is physically self-consistent: neglecting 
correlations outside the sufficiently large extended molecule implies that they are 
significant only inside it, 
$\boldsymbol{\hat{\Sigma}}_{corr} \simeq \boldsymbol{\hat{\Sigma}}_{i,corr}$,
i.~e., the parenthesis of Eq.~(\ref{eq-dyson-corr}) also vanishes. 
The \emph{mathematical} equivalence of these two physical conditions is guaranteed by 
Eqs.~(\ref{eq-dyson-corr}) and (\ref{eq-dyson-X}), which show that the two parentheses
vanish simultaneously, and yield a unique equation in the case of a sufficiently 
large extended molecule.
Ultimately, the above physical assumption of a sufficiently large 
extended molecule is justified by 
just the problem that one wants to describe. If the opposite were true, the separation in 
uncorrelated left and right electrodes, and an extended (possibly strongly) correlated
molecule would be impossible, and the problem would be 
no longer that of the transport through a single molecule.

Practically, it is important that the extended molecule be not too large, 
such that accurate ab initio quantum-chemical calculations to correlated molecules 
of sizes $\mathcal{L}_{c}$ be feasible. As any approximation, the 
present approach also has a limited applicability. 
Importantly is however that --- unlike other approximate 
schemes justified mathematically rather than physically --- 
its limitation is physically transparent. 
The present approach can\emph{not} reliably tackle situations characterized by 
too large correlation lengths $\mathcal{L}_{c}$. Fortunately, such cases are rare. 
We can only mention a single case:
the Kondo effect, extensively studied from  
various persepctives (e.~g., Refs.~\onlinecite{Ng:88,Glazman:88,Baldea:2008b,Wegewijs:09,Baldea:2010d} 
and citations therein). As it is widely accepted, the Kondo anomaly in the zero-bias conductance is 
due to electron correlations related to coherent spin fluctuations within the Kondo cloud 
of linear size $\mathcal{L}_K \sim \hbar v_F/(k_B T_K)$. 
For typical Kondo temperatures $T_K \sim 1$\,mK$-1$\,K
and $v_F = 1.4 \times 10^6$\,m/s (Fermi velocity in gold electrodes), $\mathcal{L}_K \sim 10^3 - 10^6$\,nm,
a size which is well beyond the possibilities of any nontrivial 
ab initio study. As discussed recently 
\cite{Baldea:2010d}, Kondo-related 
properties obtained by treating exactly clusters of (much) smaller sizes,
$\mathcal{L}_{c} \ll \mathcal{L}_K$, should be considered with very special care.

We do not attempt to give a general estimate of the size $\mathcal{L}_{c}$,
a hardly feasible task for real systems. 
Instead, we shall benchmark the approach based on a sufficiently large extende molecule 
for a non-trivial correlated model. On this basis, we argue that this approach 
can provide results, which are relevant for the transport in correlated molecular systems.

So, if the extended molecule is sufficiently large, one can 
reliably determine the matrix Green's function $\boldsymbol{\hat{G}}$ 
of the embedded molecule as
\begin{eqnarray}
\displaystyle
\label{eq-dyson-ab-initio}
\boldsymbol{\hat{G}_{}^{-1}} & \simeq & 
\boldsymbol{\hat{G}_{i,0}^{-1}} - \boldsymbol{\hat{\Sigma}_{i, corr}} -  
\boldsymbol{\hat{\Sigma}_{0,L}} - \boldsymbol{\hat{\Sigma}_{0,R}} \ , \\
\boldsymbol{\hat{G}_{}^{-1}} & \simeq & 
\boldsymbol{\hat{G}_{i}^{-1}} - \boldsymbol{\hat{\Sigma}_{0,L}} - \boldsymbol{\hat{\Sigma}_{0,R}} \ ,
\label{eq-dyson-work}
\end{eqnarray}
provided that one is able to accurately compute the matrix Green's function
$\boldsymbol{\hat{G}_{i}}$ of an isolated, but still sufficiently large extended molecule. 
In view of the physical motivation behind it, we shall refer to 
the approximation involved in Eqs.~(\ref{eq-dyson-ab-initio}) and (\ref{eq-dyson-work})
as the approximation of a sufficiently large extended molecule.
This concept plays a central role for all the 
subsequent considerations of the present paper.

According to Eq.~(\ref{eq-dyson-work}), to compute 
the full matrix Green's function $\boldsymbol{\hat{G}}$ of a sufficiently large extended molecule,  
in addition to $\boldsymbol{\hat{G}_{i}}$, one only needs $\boldsymbol{\Gamma_{x}}$ and 
${\boldsymbol{\hat{\Sigma}}_{0,x}}$ ($x=L,R$). More precisely, what one only needs is the retarded 
self-energy $\boldsymbol{\Sigma_{0, x}^r}$, because all the other quantities  
can be expressed in terms of it as follows \cite{Meir:92,Ferretti:05}
\begin{eqnarray}
\boldsymbol{\Gamma_{x}} & = & 
i \left[\boldsymbol{\Sigma_{0, x}^r} - \left(\boldsymbol{\Sigma_{0, x}^r} \right)^ {\dagger} \right] , \nonumber \\
\boldsymbol{\Sigma_{0,x}^{<}} & = & i f_{x} \boldsymbol{\Gamma_{x}}\ , \label{eq-Gammas} \\
\boldsymbol{\Sigma_{0,x}^{>}} & = & - i \left(1 - f_{x}\right) \boldsymbol{\Gamma_{x}} \ . \nonumber
\end{eqnarray}

Consequently, the hard part to calculate the current is to obtain  
the matrix Green's function $\boldsymbol{\hat{G}_{i}}$ of the isolated extended molecule. 
[As shown below, it even suffices to compute the retarded block $\boldsymbol{{G}_{i}^{r}}$, 
cf.~Eqs.~(\ref{eq-G-r-i}) and (\ref{eq-I}).] 
By inserting Eqs.~(\ref{eq-G-lesser}), (\ref{eq-G-r}), (\ref{eq-G-a}), and (\ref{eq-Gammas})
into Eq.~(\ref{eq-I-2}), one can express the steady-state current through 
a sufficiently large extended molecule as
\begin{equation}
\label{eq-I}
I \simeq \frac{e}{h} \int d\,\varepsilon \mathbf{Tr}
\left[ f_{L}(\varepsilon) - f_{R}(\varepsilon)\right]
\boldsymbol{\Gamma_{L}} (\varepsilon) \boldsymbol{G_{}^{r}} (\varepsilon) 
\boldsymbol{\Gamma_{R}} (\varepsilon) \boldsymbol{G_{}^{a}} (\varepsilon) \ .
\end{equation}
In the absence of correlations $\boldsymbol{G^{r,a}} \rightarrow \boldsymbol{G^{r,a}_{0}}$, 
Eq.~(\ref{eq-I}), which reduces to the familiar Landauer formula, is an exact result
probably first deduced in Ref.~\onlinecite{Caroli:71} and often discussed later,
e.~g. \cite{Meir:92,HaugJauho,Datta:05,Ferretti:05,Bergfield:09}.
Although similar to the uncorrelated case, 
Eq.~(\ref{eq-I}) does account for electron correlations, because 
$\boldsymbol{ \left({G^{r,a}}\right)^{-1} - \left({G_{0}^{r,a}}\right)^{-1} \simeq - \Sigma_{i,corr} \neq 0}$. 
It becomes similar to Eq.~(10) of Ref.~\onlinecite{Meir:92}, deduced for proportionate 
coupling, if one approximates $\boldsymbol{\Sigma} \simeq \boldsymbol{\Sigma_0}$ there, 
in the spirit of the present paper. 
It is also similar to Eq.~(17) of Ref.~\onlinecite{Ferretti:05};
if the correction factor used there 
is set to unity ($\boldsymbol{\Lambda} = 1$), that equation coincides with the present Eq.~(\ref{eq-I}).
The physical motivation of the coincidence of these two equations is clear: 
as seen in the definition of $\boldsymbol{\Lambda}$ [cf.~Eq.~(13) of Ref.~\onlinecite{Ferretti:05}], 
this factor accounts for the renormalization of the electrode self-energies and should approach unity
if the extended molecule becomes sufficiently large \cite{Lambda-at-Fermi-energy}.

Eq.~(\ref{eq-I}) seems odd for the correlated case, as it only contains the retarded Green's function 
[notice that $\boldsymbol{G^{a}} = (\boldsymbol{G^{r}})^{\dagger}$] but not the lesser Green's function 
[cf.~Eq.~(\ref{eq-I-2})]. However there is no contradiction. 
$\boldsymbol{G^{<}}$ needed in Eq.~(\ref{eq-I-2}), 
can be expressed in terms of $\boldsymbol{G^{r}}$ 
only if the \emph{bare} embedding self-energies $\boldsymbol{\Sigma}_{x,0}$ of 
Eq.~(\ref{eq-Gammas}) are used instead 
of the dressed ones  $\boldsymbol{\Sigma}_{x}$ in Eq.~(\ref{eq-G-lesser}), and this is legitimate 
\emph{only} for a sufficiently large extended molecule. 
For smaller sizes, both (nonequilibrium) 
$\boldsymbol{G^{r}}$ and $\boldsymbol{G^{<}}$ are needed to 
compute the current. 

The retarded Green's function block $\boldsymbol{G^{r}}$ of the sufficiently large 
extended molecule coupled to electrodes can be obtained from that 
of the isolated one $\boldsymbol{G^{r}_{i}}$ from a Dyson equation similar to Eq.~(\ref{eq-G-r,a})
\begin{equation}
\displaystyle
\label{eq-G-r-i}
\left(\boldsymbol{G_{}^{r}}\right)^{-1} \simeq \left(\boldsymbol{G_{i}^{r}}\right)^{-1} - 
\boldsymbol{\Sigma_{0,L}^{r}} - \boldsymbol{\Sigma_{0,R}^{r}} .
\end{equation}
In view of Eq.~(\ref{eq-G-r-i}), the retarded Green's function 
$\boldsymbol{G_{i}^{r}}$ of an isolated but sufficiently large molecular cluster represents
the crucial quantity, which is needed to compute the current from Eq.~(\ref{eq-I}).
For real molecules, $\boldsymbol{G_{i}^{r}}$ can be obtained, e.~g., within 
diagrammatic schemes to approximate the self-energy $\boldsymbol{\Sigma_{i,corr}^{r}}$. 
One of these schemes are the ADC-approximations widely employed for accurate electronic structure 
calculations \cite{Schirmer:83,Schirmer:84,Schirmer:89}.
Another scheme is based on the GW approximation, which was proposed long time ago for metallic 
solids \cite{Hedin:65} and used in several recent studies of small molecules
\cite{Stan:09,Thygesen:10}
and nanotransport \cite{Darancet:07,Thygesen:07,Thygesen:08b,Thygesen:08c,Millis:08,Millis:09}. 
It is certainly of interest to attempt to use such schemes for real systems,
where exact results are not available,
but applying them here in conjunction with Eqs.~(\ref{eq-dyson-ab-initio}) and (\ref{eq-I}) 
is less useful conceptually. 
They represent certain mathematical approximations,
which can be justified by comparing their predictions with experiments, but 
their validity in physically unclear.
By constrast, our approximation possesses a precise
physical content. Electrons can be strongly correlated in the molecule, but if 
interactions responsible for these correlations 
remain localized within a restricted spatial domain, 
the larger the extended molecule, the more accurate are the results; 
they should become exact in the limit of an infinite extended molecule \cite{Roberts:80,Ferretti:05}.

Our present goal is to document the approximation of a sufficiently large extended molecule,
and, to facilitate the interpretation of the results, we do not use any further approximation.
Therefore, we shall employ a simple model of a correlated molecule, which we solve exactly 
within a full CI numerical calculation. 
The exact retarded and advanced Green's functions of the isolated extended molecule will be computed 
with the aid of the Lehmann representation (case of zero temperature) \cite{mahan}
\begin{equation}
\label{eq-G-i-r}
\displaystyle
\left[ \boldsymbol{{G}_{i}^{r}} (\varepsilon)\right]_{j,k} = 
\left[ \boldsymbol{{G}_{i}^{a}} (\varepsilon)\right]_{k,j}^{\ast}  =  
\sum_{\lambda} 
\frac{\left \langle 0\right\vert a_j\left \vert \lambda\right\rangle 
\left \langle \lambda\right\vert a_{k}^{\dagger}\left \vert 0\right\rangle}
{\varepsilon - \varepsilon_{\lambda}^{EA} + i 0^{+}}
+
\sum_{\nu} 
\frac{\left \langle 0\right\vert a_j\left \vert \nu\right\rangle 
\left \langle \nu\right\vert a_{k}^{\dagger}\left \vert 0\right\rangle}
{\varepsilon + \varepsilon_{\nu}^{IP} + i 0^{+}} \ .
\end{equation}
Here, $a_j$ and $a_k^{\dagger}$ are second-quantized operators that annihilate and 
create electrons at the sites (``atomic orbitals'') $j$ and $k$ within the extended molecule,
whose ground state is denoted by $\vert 0 \rangle$, $\varepsilon_{\nu}^{IP}$  and 
$\varepsilon_{\lambda}^{EA}$ stand for the $\nu^{th}$ ionization energy and 
$\lambda^{th}$ electro-affinity, and $\{\vert \nu \rangle\}$ and $\{\vert \lambda \rangle\}$ 
represent the corresponding complete sets of eigenstates. 
\section{The model}
\label{sec-model}
To be specific, we shall consider below a one-dimensional discrete model of a two-terminal 
setup known as the 
interacting resonant level model (IRLM) \cite{Mehta:06,Bohr:07,BoulatUnitaryLimit:08,Meden:10}. 
The physical system it describes is a molecule modeled by a single level of energy $\alpha$ 
attached to the left ($L$) and right ($R$) electrodes via nearest-neighbor electron hopping 
characterized by the resonance integrals $\beta_L$ and $\beta_R$, respectively. 
The semi-infinite electrodes 
are kept to fixed chemical potentials $\mu_{L,R}$, and contain noninteracting electrons 
hopping between adjacent sites (resonance integrals $\xi_{L,R}$). 
By ignoring the electron spin, subtle coherent fluctuations (like those responsible for the 
Kondo effect \cite{Ng:88,Glazman:88}) are disregarded, but the 
Coulomb interactions $J_{L,R}$ at the molecule-electrode contacts, which are included, 
induce nontrivial electron correlations. They bring about an interesting physical behavior 
that attracted theorists' attention recently 
\cite{Mehta:06,Bohr:07,BoulatUnitaryLimit:08,Meden:10}. 
The important role played by interaction at the electrode-molecule contacts also made the object 
of recent experimental investigations \cite{Xu:07}.

To avoid problems related to the Fermi energy alignment it is helpful to keep a constant 
band filling factor; it will be chosen 1/2, which amounts to consider a number of sites twice 
the number of electrons. We shall split the total Hamiltonian $H$ of this system 
to explicitly write the part $H_M$ that specifies the extended molecule
\begin{eqnarray}  
\displaystyle
H & = & H_{L} + H_{R} + H_{LM} + H_{MR} + H_{M} \ ,  \nonumber \\
H_{L}  = & &
\mu_{L} \sum_{ l=-M_L - 1}^{-\infty} a_{l}^{\dagger} a_{l}^{}
- \xi_L \sum_{l=-M_L - 1}^{-\infty} \left(a_{l}^{\dagger} a_{l-1}^{} + h.c. \right) \ ,
\nonumber \\
H_{R} = & &
\mu_{R} \sum_{r=M_R + 1}^{+\infty}  a_{r}^{\dagger} a_{r}^{}
-\xi_R \sum_{r=M_R + 1}^{+\infty} \left(a_{r}^{\dagger} a_{r+1}^{} + h.c. \right) \ ,
\label{eq-general-H} \\
H_{LM} & & =
 - \xi_L \left(a_{-M_L}^{\dagger} a_{-M_L - 1}^{} +  h.c.\right) \ , \nonumber \\
H_{MR} & & = - \xi_R \left (a_{M_R}^{\dagger} a_{M_R + 1}^{} + h.c. \right) \ , \nonumber 
\end{eqnarray} 
and
\begin{eqnarray}  
\displaystyle
H_{M}  = & & 
\mu_{L} \sum_{ l=-1}^{-M_L} a_{l}^{\dagger} a_{l}^{}
- \xi_L \sum_{l=-1}^{-M_L+1} \left(a_{l}^{\dagger} a_{l-1}^{} + h.c. \right)
\nonumber \\ 
& &
+ \mu_{R} \sum_{r=1}^{M_R}  a_{r}^{\dagger} a_{r}^{}
-\xi_R \sum_{r=1}^{M_R-1} \left(a_{r}^{\dagger} a_{r+1}^{} + h.c. \right) \nonumber  \\
& & - \beta_L \left(a_{-1}^{\dagger} a_{0}^{} +  h.c.\right) 
- \beta_R \left (a_{+1}^{\dagger} a_{0}^{} + h.c. \right) \ 
\label{eq-H_M} \\
& &
+  J_L \left(a_{0}^{\dagger} a_{0}^{} - \frac{1}{2}\right) \left(a_{-1}^{\dagger} a_{-1}^{} - \frac{1}{2}\right)
+ J_R \left(a_{0}^{\dagger} a_{0}^{} - \frac{1}{2}\right) \left(a_{+1}^{\dagger} a_{+1}^{} - \frac{1}{2}\right) 
+ \alpha  a_{0}^{\dagger} a_{0}^{} \ .
\nonumber
\end{eqnarray} 
In the above equation, the extended molecule comprises, besides the ``molecule'' itself,
represented by a single site (labeled by $0$), parts of the left and right electrodes 
(sites $-M_L, \ldots, -1$ and $1, \ldots, M_R$, respectively). Importantly,
the energy of the molecular level $\alpha$ can be tuned by direct molecular orbital gating,
as demonstrated by a remarkable very recent experiment \cite{Reed:09}, or 
electrochemical gating \cite{KuznetsovElectrocheGating:00,Cronin:04,Albrecht:07,Davies:08}
in a way that is similar to the usage of a ``plunger'' gate electrode in 
quantum-dot nanoelectronic devices \cite{Goldhaber-GordonNature:98,Wiel:00}.
Unlike in quantum-dot based single-electron transistors, 
where the relevant $\alpha$-range is of the order of a few meV
\cite{Goldhaber-GordonNature:98,Wiel:00}, in the aforementioned experiment \cite{Reed:09}
the energy of the molecular level $\alpha$ 
could be varied, remarkably, within much broader ranges: 
$1.1\,eV \alt \varepsilon_F - \alpha \alt 1.9\,eV$ for 
single-electron transistors based on 1,8-octanedithiol and 
$0.4\,eV \alt \varepsilon_F - \alpha \alt 1.8\,eV$ for those based on 1,4-benzenedithiole
by varying the gate potential $V_G$ within $ -3.3\,V < V_G < 0 $ and $ -3\,V < V_G < 3\,V $,
respectively. In the presently considered case of 
half filling, the number of sites of the extended molecule $N=M_L + M_R + 1$ 
must be even. This implies that $M_L$ is odd and $M_R$ is even or vice versa, and therefore 
it is impossible to consider perfectly symmetrical extended molecules. The numerical 
results presented below correspond to the most symmetrically possible situation 
$M_L=M_R \pm 1 $, but the conclusions of the present paper are not qualitatively 
affected by this choice.

For one-dimensional models with nearest-neighbor hopping (the class to which the 
above model belongs), the contact ``surfaces'' between the extended molecule and the left 
and right electrodes reduce to two points, which are located at $-M_L$ and $+M_R$, respectively.
Therefore, the self-energies $\boldsymbol{\Sigma_{0,L}^{r}}$ and $\boldsymbol{\Sigma_{0,R}}^{r}$ 
needed to calculate the current of Eq.~(\ref{eq-I}) possess a single nonvanishing elements, namely 
$\left(\boldsymbol{\Sigma_{0,L}^{r}}\right)_{-M_L, -M_L} \equiv \Sigma_L$ and 
$\left(\boldsymbol{\Sigma_{0,R}^{r}}\right)_{M_R, M_R} \equiv \Sigma_R$. 

For model (\ref{eq-general-H}), $\Sigma_{x}^{r}(\varepsilon) = \beta_{x}^2 g_{x}^{r} (\varepsilon)$ 
where $g_{x}^{r}$ is the retarded Green's function at the ends of semi-infinite electrodes coupled 
to the extended molecule. 
By using the expression of $g_{x}^{r}(\varepsilon)$ \cite{Datta:05},
one straightforwardly obtains
\begin{equation}
\displaystyle
\Sigma_x (\varepsilon) =  \frac{\beta_x^2}{2 \xi_x^2}(\varepsilon - \mu_x) - \frac{i}{2} \Gamma_x(\varepsilon) \ ,
\end{equation}
\begin{equation}
\displaystyle
\Gamma_x(\varepsilon) = \frac{\beta_x^2}{\xi_x^2}
\sqrt{4 \xi_x^2 - (\varepsilon - \mu_x)^2} \ \ \theta(2 \xi_x - \vert \varepsilon - \mu_x \vert) \ ,
\end{equation}
where $\theta$ is the step (Heaviside) function. 
The above expressions, along with the Dyson equation for the retarded Green's function 
deduced from Eq.~(\ref{eq-dyson-work})
\begin{equation}
\label{eq-G-r-dyson}
\displaystyle
{G}_{j,k}^{r} (\varepsilon) \simeq
\left[ \boldsymbol{{G}_{i}^{r}} (\varepsilon)\right]_{j,k} + 
\left[ \boldsymbol{{G}_{i}^{r}} (\varepsilon)\right]_{j,-M_L} 
\Sigma_L (\varepsilon)
{G}_{-M_L, k}^{r} (\varepsilon) + 
\left[ \boldsymbol{{G}_{i}^{r}} (\varepsilon)\right]_{j,M_R} 
\Sigma_R (\varepsilon)
{G}_{M_R, k}^{r} (\varepsilon) \ ,
\end{equation}
complete the formulae needed to compute the current (\ref{eq-I}).
\section{Numerical results for the linear conductance}
\label{sec-conductance}
We note that, although Eq.~(\ref{eq-I}) as well as the 
subsequent equations for the nonequilibrium Green's functions are valid for an arbitrary voltage, 
in the presentation of the numerical results of this section
we shall restrict ourselves to the case of linear conductance 
$G \equiv d I/d V\vert_{V=0}$. For completeness,
we give its expression for the case of a sufficiently 
large extended molecule, which can be obtained by taking
$\mu_{L,R} = \varepsilon_F \pm eV/2$ and $V\to 0$ in Eq.~(\ref{eq-I})
\begin{equation}
\displaystyle
\label{eq-conductance}
G /G_0  \simeq  
\Gamma_{L}(\varepsilon_F) 
\vert G_{-M_L, M_R}^{r} (\varepsilon_F)\vert^2 
\Gamma_{R}(\varepsilon_F) \ ,
\end{equation}
where $G_0 \equiv e^2/h$ is the conductance quantum. For the sake of simplicity, we shall 
suppose that the electrodes are identical and therefore
$\xi_L = \xi_R \equiv \xi$, $\beta_L = \beta_R \equiv \beta$, 
$\Gamma_L (\varepsilon_F) = \Gamma_R (\varepsilon_F) = 2 \beta^2/\xi \equiv \Gamma$, 
and $J_L = J_R \equiv J$. Unlike in most of the studies on the IRLM or similar models 
(see, e.~g., Refs.~\onlinecite{Mehta:06,Bohr:07,BoulatUnitaryLimit:08,Baldea:2008b,Baldea:2009a}), 
where $\xi$ is chosen 
as the unit of energy, we shall use throughout $\beta = 1$. The reason is that we aim to compare
the cases of organic and metallic electrodes among themselves, 
and their bandwidths ($4\xi$ within our model) 
significantly differ. On the other side, we do not expect substantial differences 
between the resonance integrals $\beta$ corresponding to a molecule attached to organic and metallic 
electrodes, which should be of the order of a few electron volts. Basically, the same can also 
be expected for the Coulomb contact interaction $J$.

Concerning the size $N$ of the extended molecule we note the following. The main interest
of the present paper is not the IRLM itself. Rather, we aim to unravel what one 
can learn from the results for this model for theoretical approaches to electron 
transport in real systems based on feasible ab initio quantum-chemical calculations able 
to accurately 
treat strong electron correlations. With the presently available computational 
resources it is unlikely that accurate calculations can include more than a few electrode layers. 
Within the present model, this would correspond to at most $N\approx 12$.
Therefore, it would make little sense to examine the results 
for the IRLM up to the largest size ($N\approx 26$), which can be tackled within full CI calculations 
by calculating the Green's function of the isolated molecule $\boldsymbol{G_{i}^{r}}$ 
from Eq.~(\ref{eq-G-i-r}) based on the Lanczos algorithm 
\cite{weikert:7122,Baldea:97,Baldea:2008,Baldea:2009b,Baldea:2009a,Baldea:2010a,Baldea:2010d},
nor to carry out a finite scaling analysis of the numerical results to extract 
the limit $N\to \infty$. 
For the sizes of interest,
the straightforward exact numerical diagonalization can be and has been employed to compute the 
exact Green's function $\boldsymbol{G}_{i}^{r}$, Eq.~(\ref{eq-G-i-r}), within full CI calculations. 
\subsection{Case of organic electrodes}
\label{sec-organic}
To model a (chemisorbed) molecule linked to organic electrodes, one should consider that the
resonance integrals $\xi$ and $\beta$ are of comparable magnitudes, and this results in large values 
of the hybridization parameter $\Gamma=2\beta^2/\xi$. 
For illustration, in this subsection we shall examine the case where they are equal, $\xi=\beta(=1)$.

First, we want to benchmark the approximation of the sufficiently large extended molecule 
presented in Sect.~\ref{sec-theory}
against a well established exact result, namely the case where 
the molecular level is at resonance with the electrode Fermi energy 
($\alpha = \varepsilon_F$). There, similar to the cases of 
the noninteracting resonant level ($J\equiv 0$) and of the single-electron transistor 
\cite{Ng:88,Glazman:88,Wiel:00}, the conductance of the IRLM reaches the unitary limit
$G=G_0$. Physically, this results from the fact that (irrespective of the values 
of $\xi$, $\beta$, and $J$) for $\mu_L = \alpha = \mu_R$ 
all sites are half occupied, $n_j \equiv \langle 0\vert a_j^{\dagger} a_j\vert 0\rangle = 1/2$, 
and this yields, via Friedel's rule $G=G_0\sin^2(\pi n_0)$ \cite{BoulatUnitaryLimit:08}, 
a perfect transmission ($G=G_0$). Numerically, the result $G=G_0$ on resonance was obtained by means of 
time-dependent density matrix renormalization group calculations \cite{Bohr:07}. 

\begin{figure}[htb]
\centerline{
\includegraphics[width=0.25\textwidth,angle=-90]{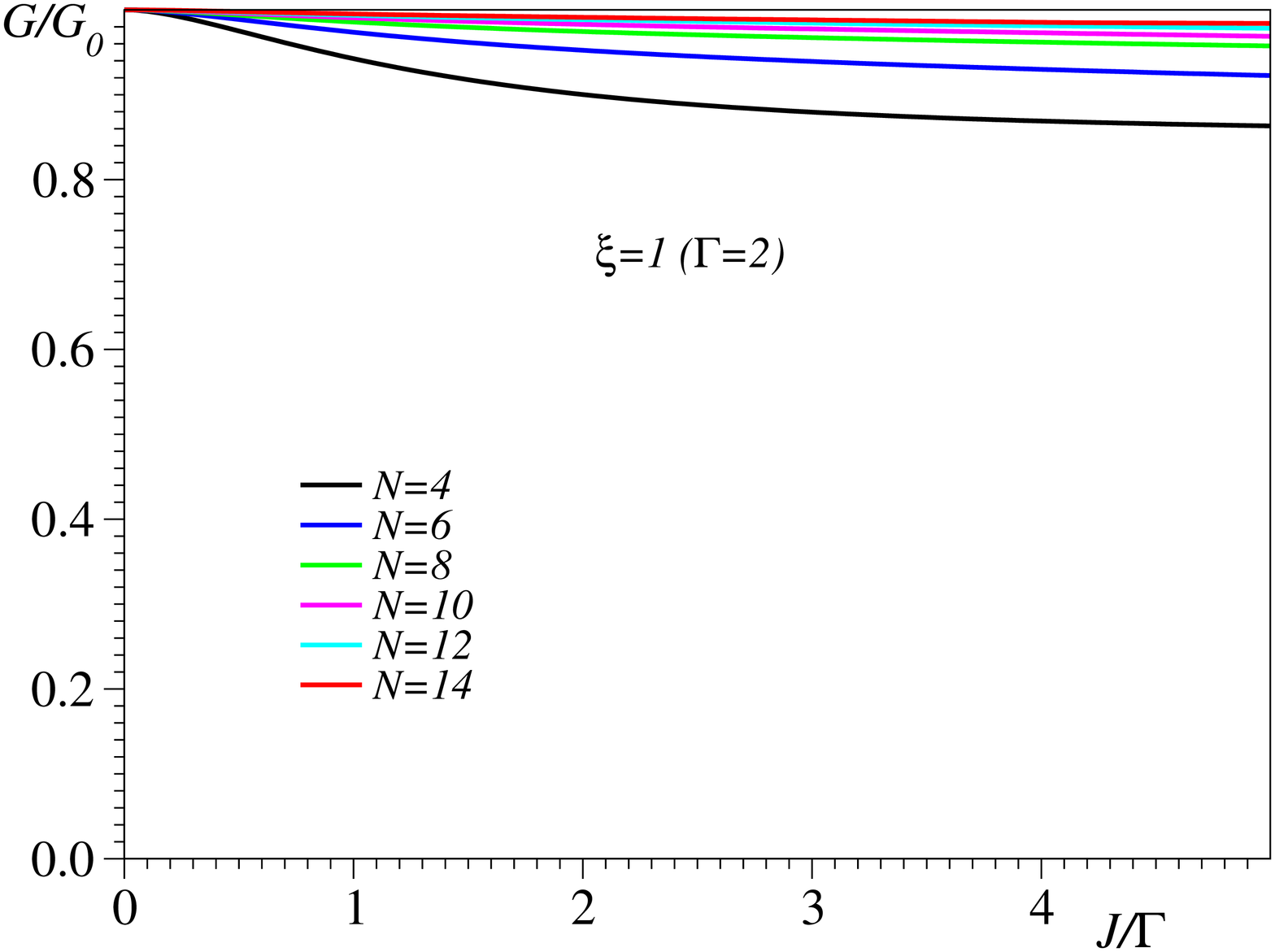}
\hspace*{10ex}
\includegraphics[width=0.25\textwidth,angle=-90]{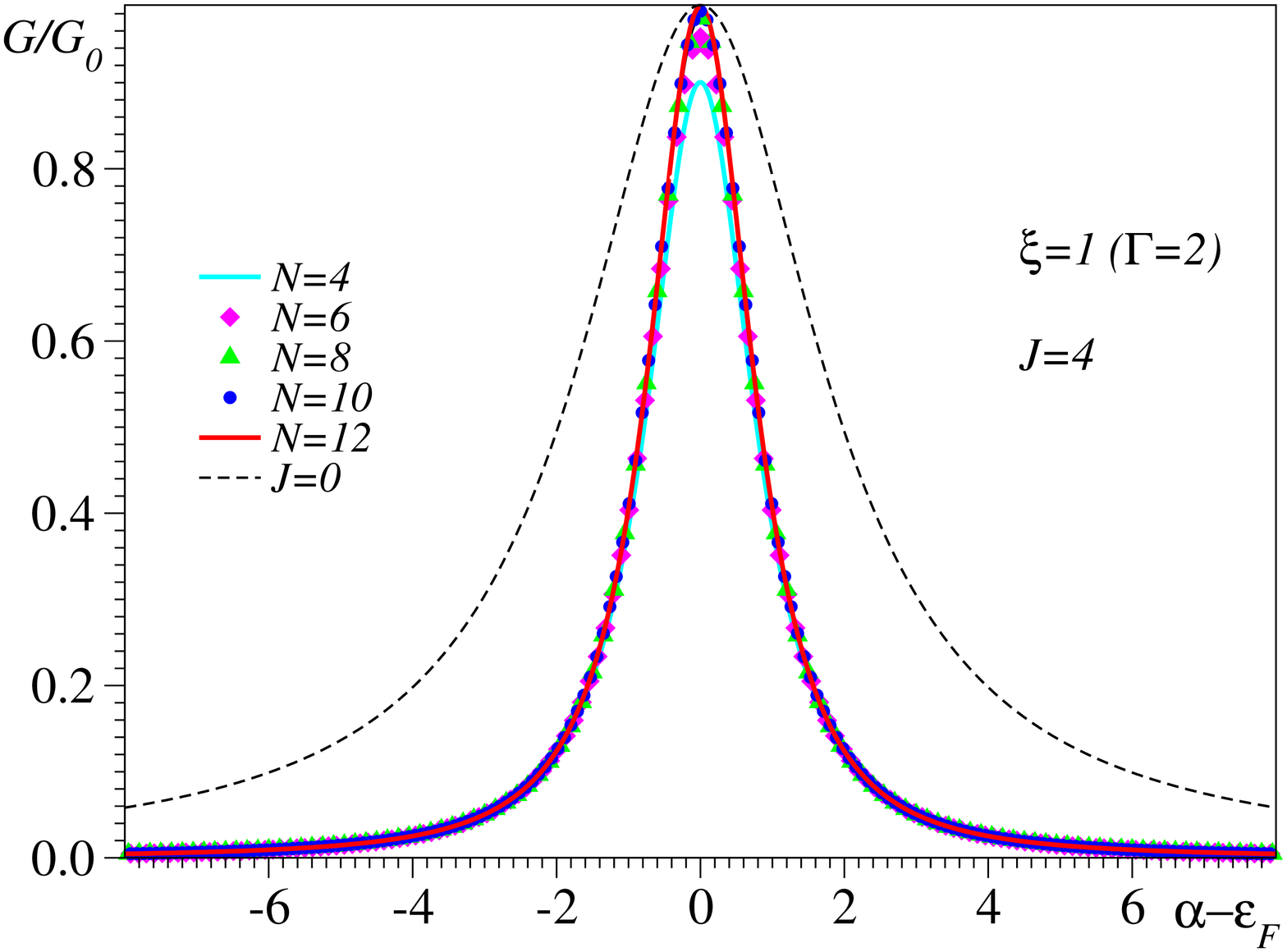}}
\caption{\label{fig:td=1} 
Conductance $G$ computed numerically 
from Eq.~(\ref{eq-conductance}) for several sizes $N$ in units of quantum conductance 
$G_0 = e^2/h$ plotted versus: (a) Coulomb contact repulsion $J$ at resonance 
($\alpha = \varepsilon_F$), and (b) molecular level energy $\alpha$ for $\xi=1$, $J=4$. 
The sizes $N$ of the extended molecules are given in the legend. The thin dashed line 
in panel (b) corresponds to a vanishing Coulomb interaction ($J=0$).}
\end{figure}

Our results for the on-resonance case at $\beta=\xi$ are presented in 
Fig.~\ref{fig:td=1}a. 
As visible there, even the smallest possible 
extended molecule ($N=4$) represents a ``sufficiently large'' extended molecule, enabling 
to accurately reproduce the exact result.
Indeed, even for the strongest coupling shown in Fig.~\ref{fig:td=1}a 
($J=10$, which is in fact unrealistically strong),
the numerical results deviate from the exact value $G=G_0$ by $\simeq 13\%$. 
As expected from the physical analysis backing the approximation of a sufficiently large 
extended molecule, Fig.~\ref{fig:td=1} reveals that the deviation from 
the exact result diminishes with increasing sizes; for the next smallest size
($N=6$), the error is at most $5\%$. 

Because $G=G_0$ holds on resonance even for the uncorrelated case ($J=0$), where 
the results are anyway independent on the size of the extended molecule (cf.~Sect.~I), 
one might suspect that the very weak size-independence displayed in Fig.~\ref{fig:td=1}a
does not demonstrate the validity of the approximation of a 
sufficiently large extended molecule, but rather that the electron 
correlations themselves are weak. To illustrate that this is not the case, we present 
in Fig.~\ref{fig:td=1}b results for the conductance out of resonance 
($\alpha \neq \varepsilon_F$) and realistic, moderately strong Coulomb strength 
$J=2\Gamma = 4$ (and equal to the electrode bandwidth $4\xi = 4$). 
They confirm the above finding, namely  
that the size-independence is indeed very weak. In addition, the conductance for $J=4$
can be compared with that for the 
noninteracting level ($J=0$) (thin dashed line in Fig.~\ref{fig:td=1}b); the substantial 
difference between them indicates a significant electron correlation effect on the electric 
transport. 
To exemplify, we note that this Coulomb interaction diminishes the half-width of the curve 
$G=G(\alpha)$ roughly by a factor two (from $\Gamma=2$ to $0.8$).
At $\alpha=\varepsilon_F \pm \Gamma$ 
($\Gamma=2\beta^2/\xi=2$), i.~e. at the half maximum in the uncorrelated case, 
the electron correlations suppress the conductance by a factor $\sim 5$. Importantly,
for situations even farther away from resonance (and this is the usual case, including
also that of the recent 
experiments of Ref.~\onlinecite{Reed:09}), Fig.~\ref{fig:td=1}b shows that the 
suppression factor becomes considerably larger.

On the basis of the results presented in this subsection, we claim that 
the correlated electron transport through a molecule attached to organic electrodes
can be accurately obtained by carrying out microscopic calculations for small  
extended molecules; besides the (true) molecule of interest, they should only include a 
reduced number of electrode layers.
\subsection{Case of metallic electrodes}
\label{sec-metal}
In comparison with organic electrodes, metallic electrodes are characterized by a 
larger bandwidth ($4\xi$). To model (physisorbed) molecules weakly coupled to 
metallic electrodes within the framework of the present paper, one should consider 
values of $\xi$ larger than in Sect.~\ref{sec-organic}. 

\begin{figure}[htb]
\centerline{\includegraphics[width=0.25\textwidth,angle=-90]{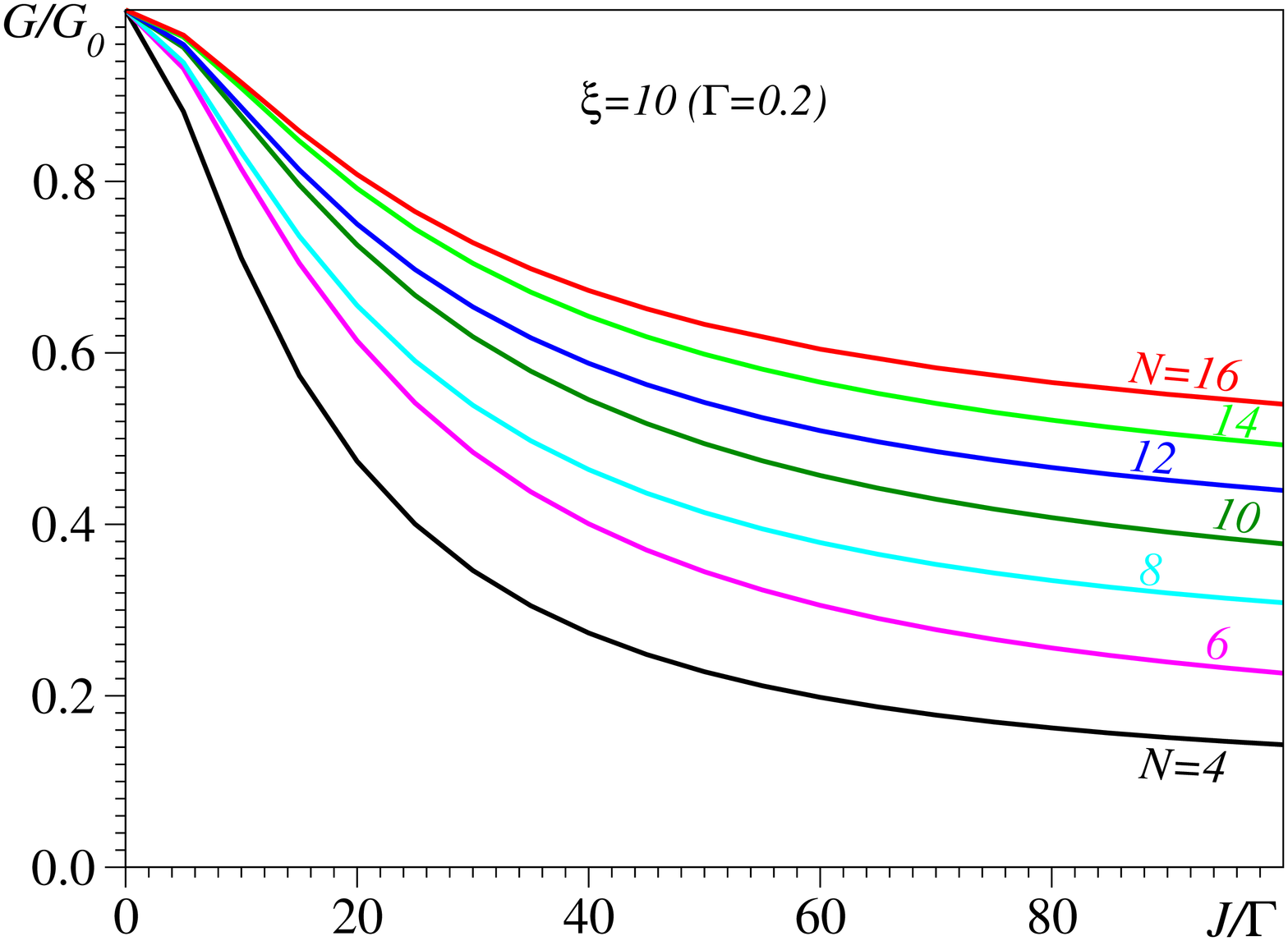}
\hspace*{10ex}
\includegraphics[width=0.25\textwidth,angle=-90]{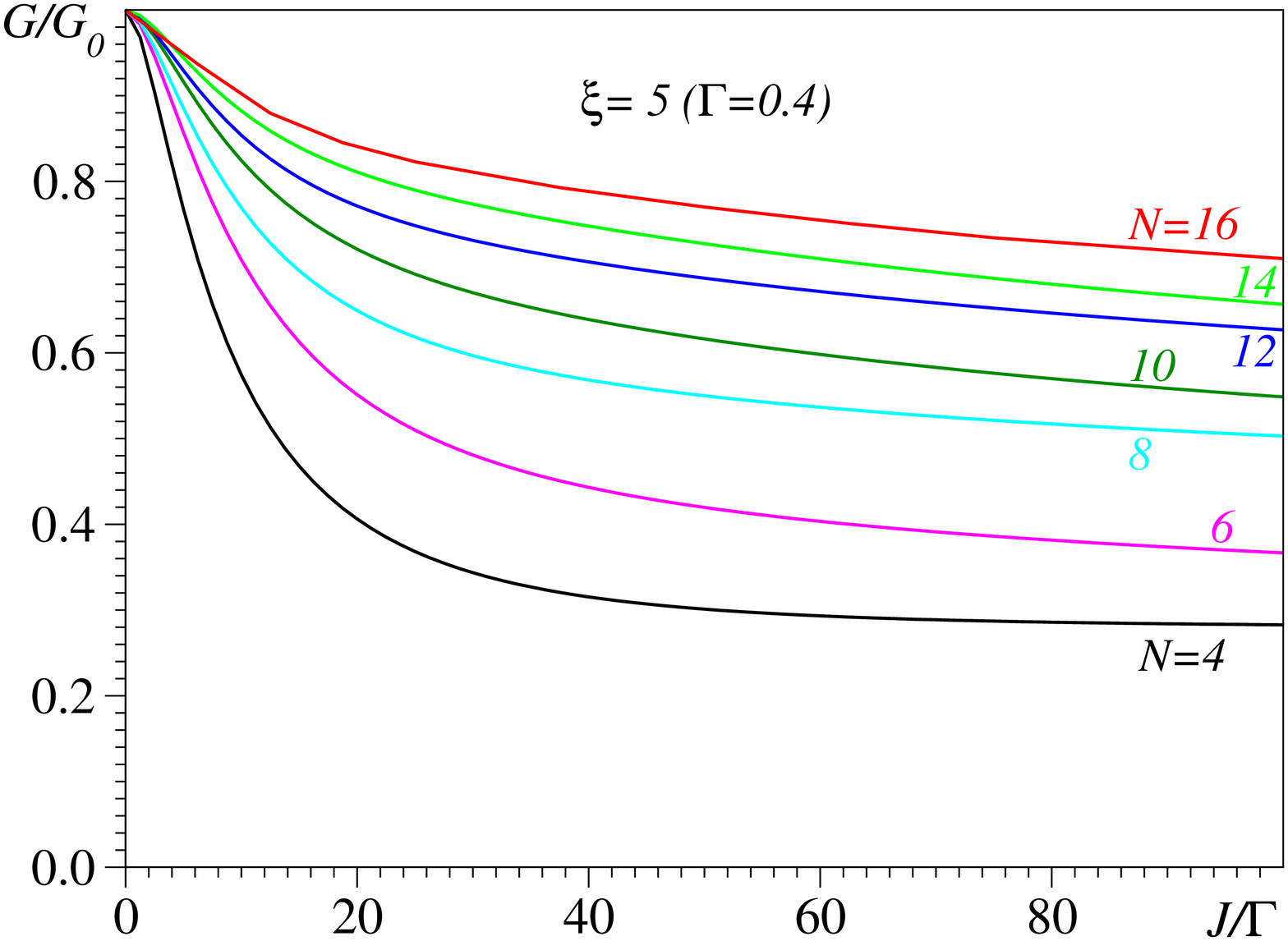}}
\caption{\label{fig:unitLim} 
Conductance $G$ in units of quantum conductance 
$G_0 = e^2/h$ plotted versus contact interaction $J$ at resonance $\alpha = \varepsilon_F$ for 
$\xi=10$ and $\xi=5$ computed using extended molecules with 
$N=4, 6, 8, 10, 12, 14, 16$ sites (values increase upwards).}
\end{figure}

In Fig.~\ref{fig:unitLim}, we present numerical results on resonance 
for two values of the molecule-electrode resonance integral, $\xi = 10$ ($\Gamma = 0.2$) 
and $\xi = 5$ ($\Gamma = 0.4$), and several sizes $N$ of the extended molecule.
The trend that, as the extended molecule becomes larger, the linear conductance 
approaches the unitary limit ($G \to G_0$) can be seen in both panels of 
Fig.~\ref{fig:unitLim}. So, the basic physical argument behind the approximation of a 
sufficiently large extended molecule is also confirmed by these numerical results. 
However, unlike in Fig.~\ref{fig:td=1}a, the deviations of the numerical values 
of $G$ from the exact value $G_0$ visible in Fig.~\ref{fig:unitLim}a and b are significant 
and deserve further analysis.

Figs.~\ref{fig:unitLim}a and b show that 
the departure from the unitary limit is larger for a stronger Coulomb contact strength $J$. 
This behavior can be easily understood. The stronger the electron-electron interaction, 
the more important is the renormalization of the contact self-energies $\boldsymbol{\Sigma_{x}}$
from the bare self-energies $\boldsymbol{\Sigma_{0,x}}$. This renders more significant
the quantity in the parenthesis of the r.h.s.~of Eq.~(\ref{eq-dyson-X}), which was neglected 
to derive the approximate Eqs.~(\ref{eq-I}) and (\ref{eq-conductance}).
To make those parentheses negligible, one has to reduce the renormalization 
at the ends of the extended molecule due to electron correlations, 
and for this one has to sufficiently increase its size $N$.

As illustrated by Figs.~\ref{fig:unitLim}a and b, for larger $\xi$'s, the unitary limit 
can be reasonably reproduced by using the smallest possible size $N=4$ 
only for weak Coulomb interactions of the order of the bare hybridization $J \sim \Gamma$.
Unfortunately, such Coulomb strengths are of little interest. 
They are too weak, and the correlation effects 
they bring about are negligible. The counterpart of Fig.~\ref{fig:td=1}b, 
the linear conductance $G=G(\alpha)$ for $J \sim \Gamma$ is not significantly 
different from that for $J=0$.
For $\xi=10$, a Coulomb strength $J=2\Gamma=0.4$ solely yields a slight change of the 
half-width at half maximum from $\Gamma=0.2$ to 0.22,
and for $\xi=5$ and $J=2\Gamma=0.8$ the change in the half-width is only a bit larger 
(from $\Gamma=0.4$ to 0.485). 

The impact of the stronger but still realistic Coulomb contact interaction of 
Fig.~\ref{fig:td=1}b ($J=4$) is also significant at larger $\xi$. For $\xi=10$
it increases the half-width at half maximum of the curve $G=G(\alpha)$ from 
$\Gamma=0.2$ to $0.43$, whereas for $\xi=5$ the increase is from $\Gamma=0.4$ 
to $0.80$. However, the corresponding values of $J/\Gamma=20$ and $J/\Gamma=10$, 
respectively, are much larger than that of Fig.~\ref{fig:td=1}b ($J/\Gamma=2$). 
At these values of $J/\Gamma$, the curves of 
Figs.~\ref{fig:unitLim}a and b exhibit a notable $N$-dependence. However, for 
the smallest extended molecule ($N=4$) the computed conductance still amounts 
a significant fraction of the exact conductance ($\sim 45\%$ and $\sim 55\%$, respectively). 
For $N=8$, which corresponds to include three and four layers in the left and right electrodes, 
these fractions are increased to $\sim 67\%$ and $\sim 80\%$, respectively. 
In view of the state-of-the art in the field of molecular electronic
transport, characterized by large differences between theory and experiments, 
these values 
suggest that, although less accurate than for organic electrodes, achieving 
a semi-quantitative description based on ab initio quantum-chemical calculations 
is possible even in the case of metallic electrodes with moderately strong Coulomb 
contact repulsion.

\section{A simple renormalization scheme}
\label{sec-renorm}
As alternative to the approximation of a sufficiently large extended molecule, one can 
start from the exact Dyson equation (\ref{eq-G}) and attempt to use renormalized electrode 
self-energies $\boldsymbol{\Sigma_{x}}$ instead of $\boldsymbol{\Sigma_{0,x}}$, by devising 
certain diagrammatic approximations. We shall not pursue this line here, because 
as noted in Sect.~\ref{sec-theory}, the validity of such approximations is usually not 
transparent physically. We present instead a simple renormalization scheme, which turns out 
to work surprisingly well for the IRLM.  

Inspired by the rather general framework of the Landau Fermi liquid theory \cite{Abrikosov:1963}, 
we shall simply suppose that the aforementioned renormalization can be accounted for by means of a 
multiplicative $\varepsilon$-independent factor $F$,
$\boldsymbol{G^{r}}(\varepsilon)\vert_{exact} \simeq F \boldsymbol{G^{r}}(\varepsilon)$.
Then, the problem is to deduce this factor. For the IRLM at resonance, this can be done, because,
as already noted in Sect.~\ref{sec-organic}, the exact conductance is known, 
$G\vert_{\alpha=\varepsilon_F} = G_0$. This condition determines the needed factor at resonance 
for given values of $N$, $\xi$ and $J$ (remember that we keep $\beta=1$), 
but $F$ could dependent on $\alpha$. Because 
$G=G(\alpha)$ reaches its maximum at $\alpha=\varepsilon_F$ one can admit that, 
at least not too far away from resonance, $F$ only slightly depends on $\alpha$ 
and neglect this dependence altogether.

\begin{figure}[htb]
\centerline{\includegraphics[width=0.25\textwidth,angle=-90]{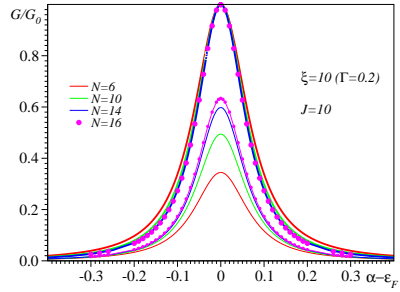}}
\caption{\label{fig:g-sc} 
Normalized conductance $G/G_0$ 
and several sizes $N$. The lower group of curves (thin lines, small points) is computed by means 
of Eq.~(\ref{eq-conductance}), and the upper group (thick lines, big points) 
are obtained within the renormalization scheme 
described in the text. Notice that, unlike the strongly size-dependent unrenormalized curves, 
the renormalized curves are weakly size-independent even 
for the rather large values $J=10$ and $\xi=10$ used, which are less favorable for the renormalized 
procedure employed.}
\end{figure}

\begin{figure}[htb]
\centerline{\includegraphics[width=0.25\textwidth,angle=-90]{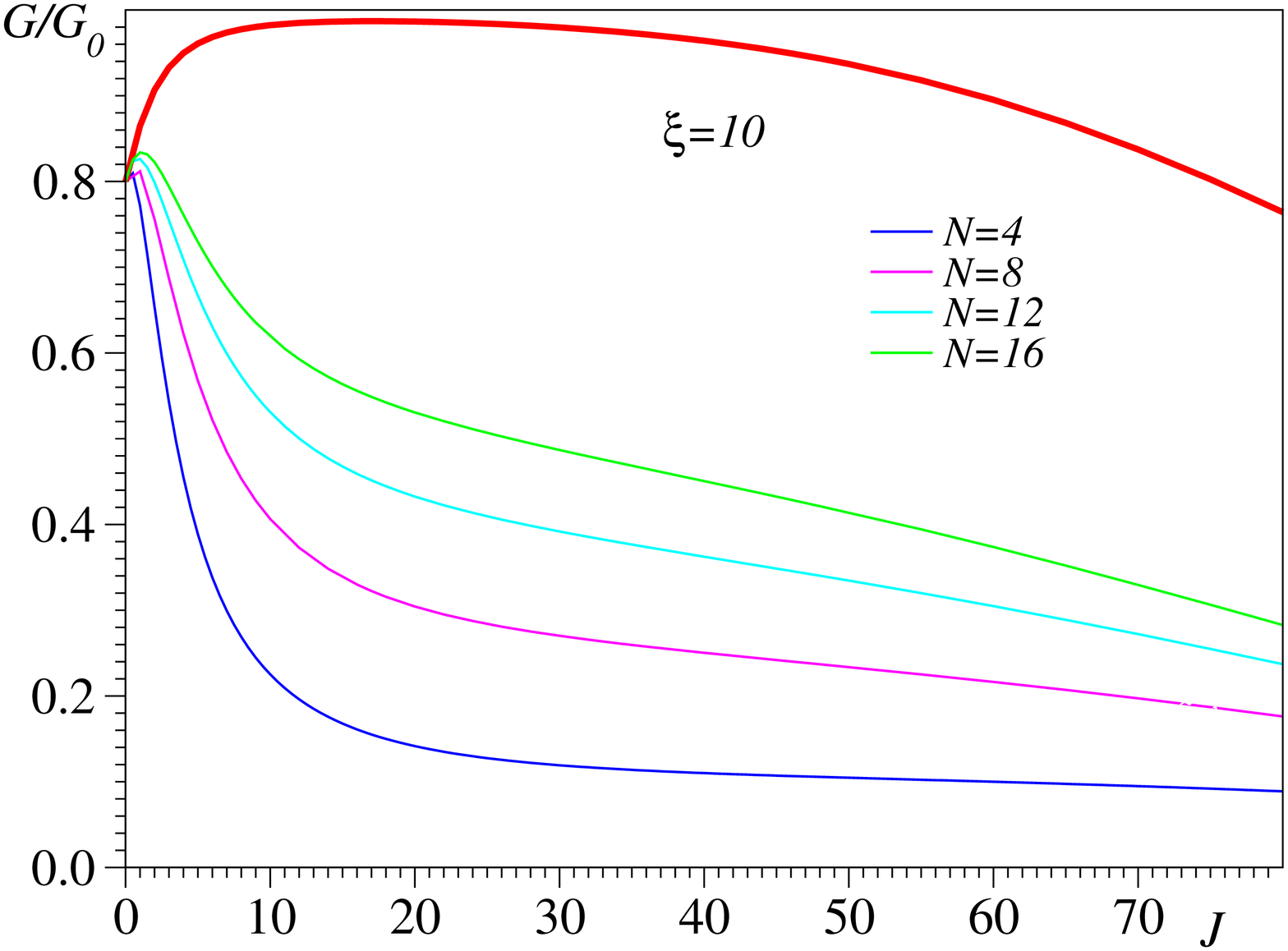}
\includegraphics[width=0.25\textwidth,angle=-90]{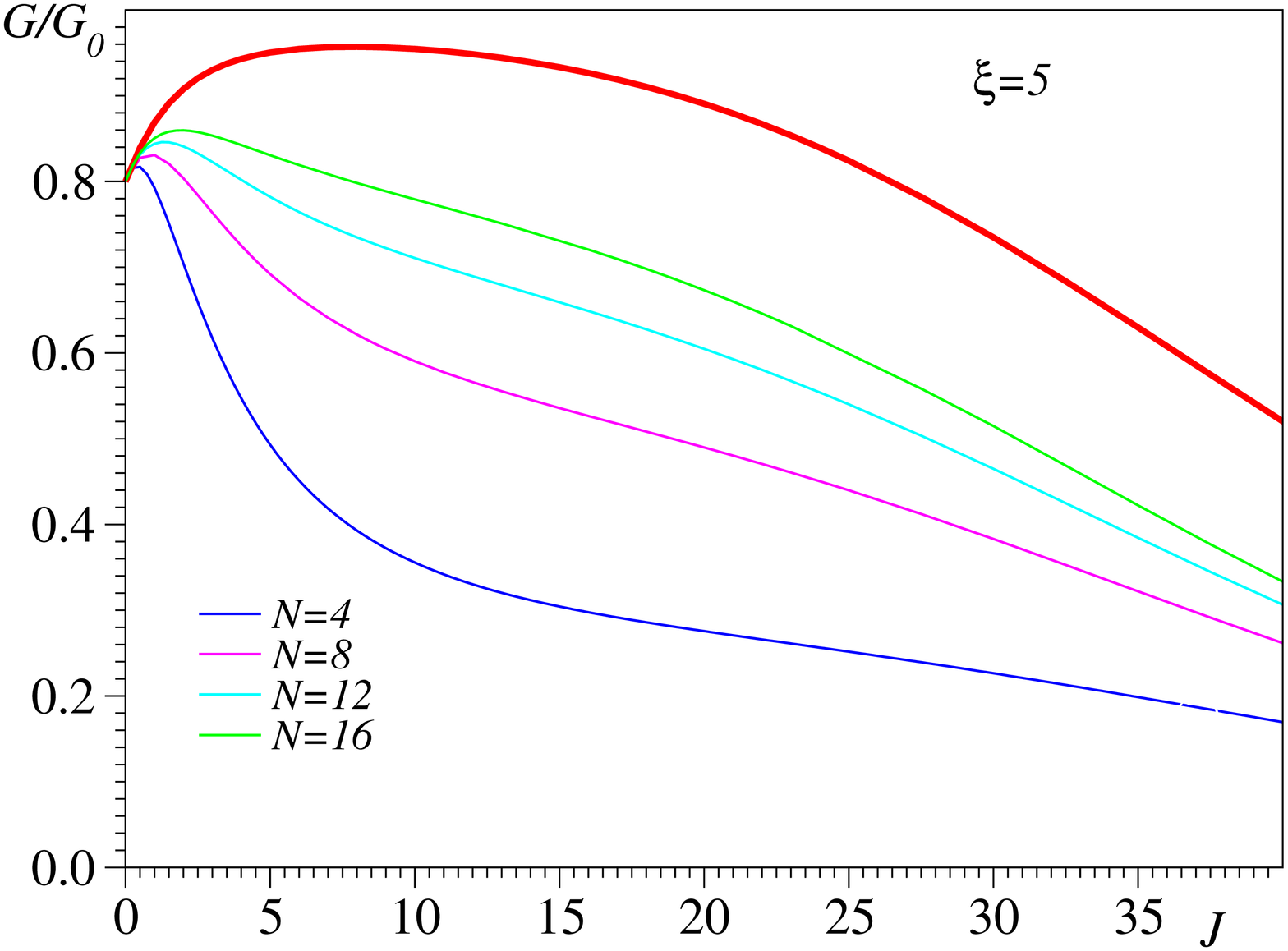}}
\centerline{\includegraphics[width=0.25\textwidth,angle=-90]{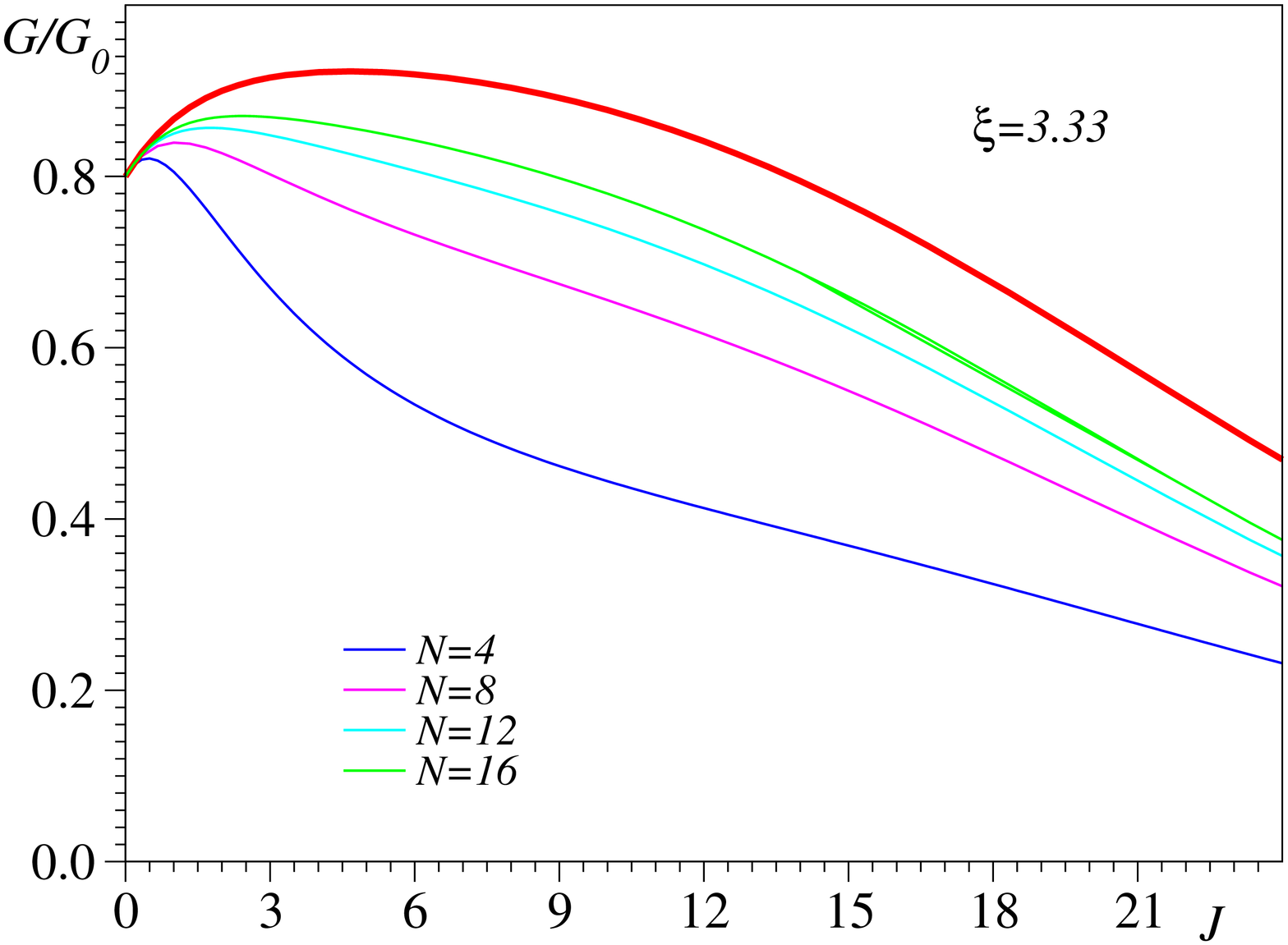}
\includegraphics[width=0.25\textwidth,angle=-90]{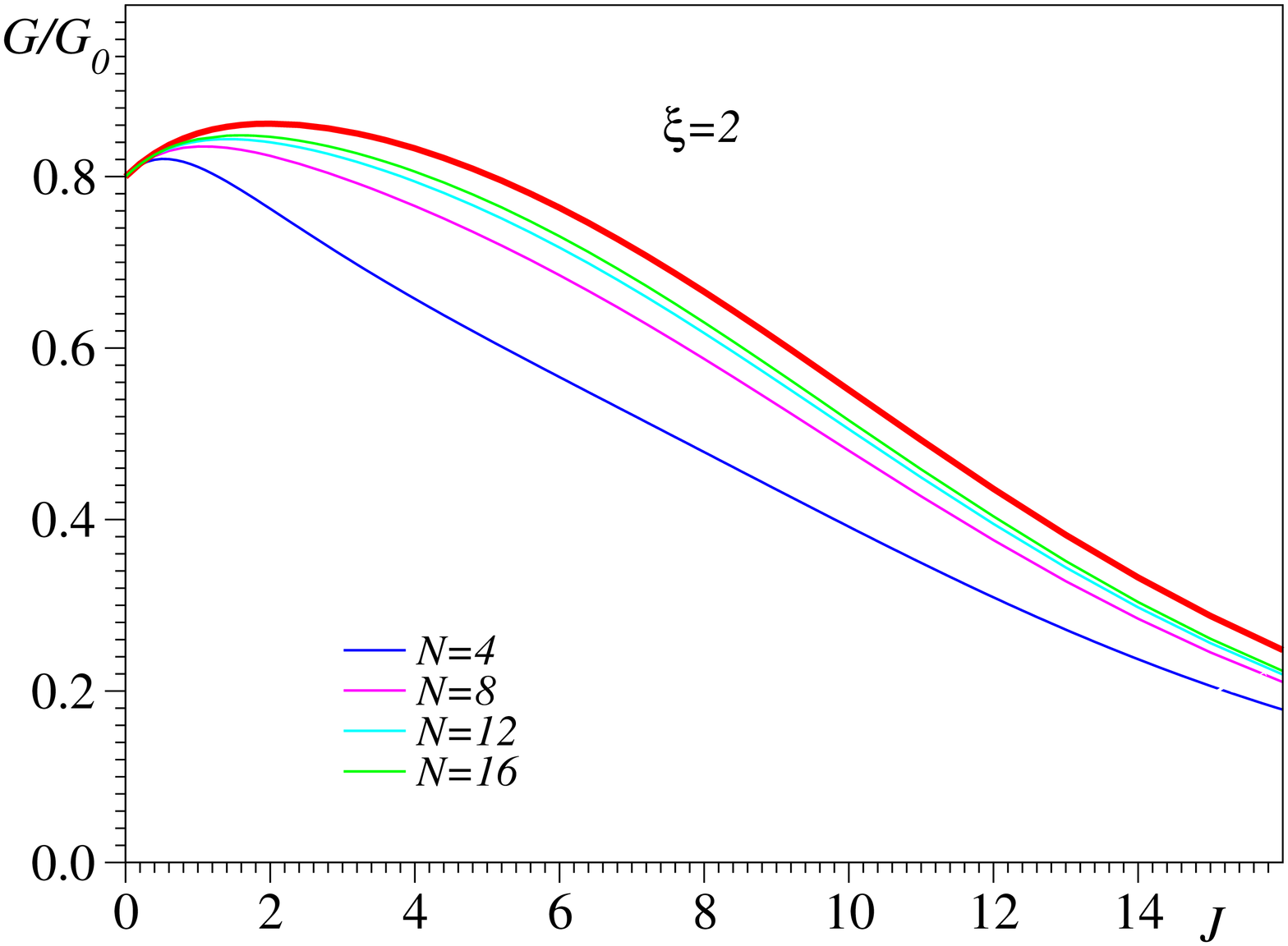}}
\caption{\label{fig:g_asym} 
Normalized conductance $G/G_0$ 
plotted versus $J/\Gamma$ for $\xi=10, 5, 3.33, 2$. All the curves were computed for the 
out-of-resonance situation $\vert \alpha - \varepsilon_F\vert = \Gamma/2$, 
which corresponds to $G/G_0 = 0.8$ at $J=0$. Notice that, the smaller the $\xi$-value, 
the stronger the size-dependence of thin curves computed from Eq.~(\ref{eq-conductance}),
and the larger the departure from the practically $N$-independent renormalized curves, 
depicted by thin lines.}
\end{figure}

We have computed and examined the curves of $G=G(\alpha)$ obtained within this renormalization 
procedure. The results are very encouraging. The renormalized curves are much less size-dependent 
than the unrenormalized ones. The smaller the values of $\xi$ and $J$, the weaker is the 
dependence on $N$. 
For illustration, we present in Fig.~\ref{fig:g-sc} for comparison 
the renormalized curves along with those unrenormalized in the rather 
unrealistic situation of very large values of $J=10$ and $\xi=10$, which corresponds 
to a very large ratio $J/\Gamma=50$.
The fact that, with increasing $N$, the renormalized curves tend to become size-independent much 
faster than the unrenormalized curves is clearly visible even in this less favorable case.
So far, we have considered the dependence on the location of molecular level $\alpha$ relative to the 
electrode Fermi energy. 
We also employed this renormalization scheme to study the $J$- and $\xi$-dependence of the 
linear conductance away from resonance $\alpha\neq \varepsilon_F$. (Remember that 
$G\vert_{\alpha=\varepsilon_F} = G_0$ for all $\xi$'s and $J$'s.) 
Results of this kind are depicted in Fig.~\ref{fig:g_asym}, which includes 
both unrenormalized and normalized curves (thin and thick lines, respectively). They confirm 
that the size-dependence is stronger at larger $\xi$'s, as already mentioned above. In addition,
they reveal the interesting fact that the dependence of the Coulomb contact interaction $J$ 
can be non-monotonic, as is the case of other previously published results \cite{Bohr:07}. 
As a general trend, we found that by increasing $J$ 
the conductance $G(J)$ gradually increases from the value $G\vert_{J=0}$, reaching its maximum 
at a Coulomb strength $J_c \sim \xi$, and decreases afterwards.
This trend can be seen by inspecting Fig.~\ref{fig:g_asym}, which also shows that the wider 
the electrode bandwidth ($4\xi$), the more pronounced is the conductance 
enhancement $G(J_c)/G(J=0)$.
\section{Discussion}
\label{sec-discussion}
%
Within the NEGF-DFT approaches, the impact of electron correlations on
the Green's function of the isolated extended molecule is accounted for approximately,
$\boldsymbol{\hat{G}_{i}}\vert_{DFT} \approx \boldsymbol{\hat{G}_{i}}$,  
by employing certain exchange-correlation (xc-)functionals. 
To determine the 
Green's function $\boldsymbol{\hat{G}}\vert_{DFT} = \left(\boldsymbol{\hat{G}_{i}^{-1}}\vert_{DFT} - 
\boldsymbol{\hat{\Sigma}_{0,L}} - \boldsymbol{\hat{\Sigma}_{0,R}}\right)^{-1}$
of the extended molecule coupled to electrodes, those approaches employ the bare
self-energies $\boldsymbol{\hat{\Sigma}_{0,L}}$ and $\boldsymbol{\hat{\Sigma}_{0,R}}$.
Because the isolated extended molecule is modeled by an effective Kohn-Sham effective Hamiltonian,
the underlying picture is a one-particle picture, and therefore the 
Landauer-type formulae (\ref{eq-I}) and (\ref{eq-conductance}) with 
$\boldsymbol{\hat{G}^{r}}$ replaced by $\boldsymbol{\hat{G}^{r}}\vert_{DFT}$ 
follow as ``exact'' results (i.~e., without any other approximation as the 
DFT-approximation itself).
  
We emphasize that Landauer-type expressions like those 
Eqs.~(\ref{eq-I}) and (\ref{eq-conductance}) are \emph{not} valid in general 
for correlated electron systems. As explained in Sect.~\ref{sec-theory}, they 
represent justified approximations in the correlated case \emph{only} when sufficiently large
extended molecules are considered.
The advantage of the approach based on this concept is its clear 
physical content. 
Eqs.~(\ref{eq-I}) and (\ref{eq-conductance}) 
hold for electron correlations that can be strong within the molecule but 
are negligible at the ends of a sufficiently large extended molecule. 
This is the reason we employed in the Dyson equation (\ref{eq-dyson-work}) the bare
self-energies $\boldsymbol{\hat{\Sigma}_{0,L}}$ and $\boldsymbol{\hat{\Sigma}_{0,R}}$, 
as in the NEGF-DFT approach. However, at variance with the latter, our approach accurately 
accounts for molecular electron correlations, which are included via the 
Green's function $\boldsymbol{G_{i}^{r}}$ of a sufficiently large extended molecule.
Because our method is based on the exact Green's function $\boldsymbol{G_{i}^{r}}$
(whose counterpart for real molecules is a Green's function computed by accurately including 
electron correlations within appropriate ab initio methods), it is clear that 
it will provide more reliable results than NEGF-DFT's. 

To give support to the intuitive physical reason underlying the concept of a sufficiently 
large extended molecule, we have presented numerical results deduced within a 
non-trivial correlated model. Based on them,
we argue that the present approach is 
able to provide results, which are interesting for the transport in 
correlated molecular systems.
To this aim, the essential prerequisite is the ability to accurately 
compute the Green's function $\boldsymbol{G}_{i}^{r}$ of the \emph{isolated} 
extended molecule. Fortunately, this can be regarded as a feasible computational task.
Such Green's functions were already obtained by means of the Lanczos algorithm 
within the ADC \cite{Schirmer:83,Schirmer:84,Schirmer:89}.
The information contained in $\boldsymbol{G}_{i}^{r}$ 
[ionization energies and electro-affinities, and pole strengths, see Eq.~(\ref{eq-G-i-r})]
can be directly compared with experimental results, which can be obtained from appropriate 
experimental investigations of the isolated molecule.
Encouragingly, accurate ADC schemes were successfully applied 
for correlated medium size molecules (e.~g., benzene), comparable to 
those of interest for molecular electronics \cite{weikert:7122}. 
Importantly, although
these methods are much more accurate than the currently considered 
GW, the computational effort is not much larger; 
it scales as $\mathcal{N}_{o}^5$ within ADC(2) and as $\approx\mathcal{N}_{o}^4$ 
within GW, where $\mathcal{N}_{o}$ is the number of atomic orbitals \cite{jochen}.

%
The difference between the present approach and the usual NEGF-approaches to 
correlated transport 
\cite{Darancet:07,Thygesen:07,Thygesen:08b,Thygesen:08c,Millis:08,Millis:09,Bergfield:09}
can be discussed by inspecting the parenthesis of the r.h.s.~of Eq.~(\ref{eq-dyson-corr}). 
In our case, this parenthesis becomes negligibly, and the difficulty to acount for 
correlations is to accurately tackle a sufficiently large extended molecule.
If we considered the true (small) molecule and neglected that parenthesis, 
we would make an approximation similar to the so-called elastic cotunneling 
approximation \cite{Averin:90,Groshev:91,Francheschi:01}. Then, the current 
would be (rather poorly) approximated by Eq.~(\ref{eq-I}) \cite{Bergfield:09}.
This amounts to take $N=3$ \cite{N=1=3}, which as seen by inspecting 
Figs.~\ref{fig:unitLim}, \ref{fig:g-sc}, and \ref{fig:g_asym}, 
is not a good approximation. Our approach goes beyond the elastic cotunneling 
approximation,
it accounts for all higher-order processes within the  
\emph{extended} molecule.
Other NEGF-approaches to correlated transport 
\cite{HaugJauho,Darancet:07,Thygesen:07,Thygesen:08b,Thygesen:08c,Millis:08,Millis:09,Bergfield:09} 
(which are also beyond the elastic cotunneling) adopt a different 
standpoint: they consider the true (small) molecule linked to 
electrodes and the evaluate the aforementioned parenthesis (which does not vanish) 
by certain Keldysh diagrammatic approximations \cite{mahan}. Recent studies 
often employ the GW approximation
\cite{Darancet:07,Thygesen:07,Thygesen:08b,Thygesen:08c,Millis:08,Millis:09}. 
Results obtained within the GW approximation
are certainly of interest \cite{Darancet:07,Thygesen:07,Thygesen:08b,Thygesen:08c}, 
although limitations and artefacts are also significant \cite{Millis:08,Millis:09}. 
Methodologically, the weak point of this approach is the fact that its physical content 
is hard to understand. Thus, it is difficult to assess a priori its validity for 
a real case. The attempt to benchmark the GW approximation against exact results 
\emph{for transport} led to ambiguous conclusions \cite{Millis:08,Millis:09}.
Within this framework, both important effects (i.~e., intramolecular correlations and 
molecule-electrode coupling) are treated in an approximate manner. Consequently, it is difficult 
to conclude which of them is better or worse approximated. It makes more sense to benchmark 
the GW approximation for those isolated extended molecules, wherein electron correlations  
play an important role, and which are also employed in transport experiments. Recent studies 
in this direction are significant, but only considered atoms and very small 
molecules so far \cite{Stan:09,Thygesen:10}.

In Introduction, we mentioned that the results for transport should be partition 
independent. Obviously, within the present approach, where correlations in distant 
parts of electrodes are neglected, the partition independence means and is demonstrated 
by the fact that, by making the extended molecule larger than a certain size 
$\mathcal{L}_c$, the results do not (notable) change.
Regarding the size-(in)dependence of the results of other approaches to correlated 
molecular transport, attention should be paid to the following important aspect. 
The fact that calculations accounting approximately for electron correlations 
(like DFT's or GW's) yield results that saturate with increasing $N$, does not 
\emph{necessarily} imply that they are physically meaningful. 
The saturation could be a mathematical artefact, because 
such approximative methods
become poorer at larger sizes and cannot reliably describe 
subtle correlations basically localized within the molecule in a system including larger and larger 
parts of \emph{uncorrelated} electrodes.
Correlation effects at larger 
sizes are likely underestimated, and this renders a correlated system resembling more an 
uncorrelated one, wherein the size-dependence is absent (cf.~Sect.~\ref{sec-introduction}),
and hence the aforementioned saturation. On the contrary, if molecular 
correlations can be accounted for by means of 
an accurate Green's function $\boldsymbol{G_{i}^{r}}$
of an isolated and sufficiently large extended molecule, size-independent results can be
trusted: they indicate that correlations are negligible at the interfaces between 
electrodes and the extended molecule.  

The present numerical results indicate that 
the extended molecules to be employed for carrying out accurate 
ab initio quantum chemical calculations need not be very large.
Particularly attractive is the case of organic conducting electrodes, for which only a few 
electrode layers have to be included (cf.~Sect.~\ref{sec-organic}). 
Another advantage is that, without metal atoms, 
ab initio calculations are less demanding. For metallic electrodes, more electrode layers 
need be included (cf.~Sect.~\ref{sec-metal}), but one can still hope to a semi-quantitative 
description. Alternatively, simpler (tight-binding) models can be utilized to include 
metallic electrode layers and their coupling to the molecule as well,
similar to the first four terms of Eq.~(\ref{eq-H_M}). 
Regarding the case of metallic electrodes, it is still worth emphasizing the following aspect. 
Experimental data indicate that electric conduction through molecules typically occurs 
by non-resonant tunneling, characterized by a rather large energy offset 
$\vert \varepsilon_{F} - \varepsilon_{HOMO}\vert$ (p-type conduction) or 
$\vert \varepsilon_{LUMO} - \varepsilon_{F}\vert$ (n-type conduction)
\cite{Choi:08,Reed:09}, which represents the counterpart of the quantity 
$\vert \alpha - \varepsilon_F\vert$ in the presently employed model.
This fact is important because, based on our calculations, we can state 
that the size dependence of the transport properties is particularly pronounced near resonance 
($\vert \alpha - \varepsilon_F\vert \approx 0$), 
but becomes much less significant away from resonance. This 
feature, which is visible if one inspects the unrenormalized curves of Fig.~\ref{fig:g-sc},
is very encouraging: one can hope to obtain quantitatively correct results 
by using extended molecules with reasonable sizes even in the case of metallic electrodes.

There is a general claim (e.~g., Ref.~\onlinecite{Paulsson:02}) 
that whether the electron transport is correlated or not solely 
depends on value of the ratio between 
the characteristic Coulomb electron-electron interaction and the hybridization 
(in our case $J/\Gamma$).
By comparing the results of Sects.~\ref{sec-organic} and \ref{sec-metal} among themselves,
we found limitations of this rule of thumb \cite{Paulsson:02}.
Based on these results, one can conclude that the ratio $J/\Gamma$ alone is not 
sufficient to distinguish between significant and insignificant correlation effects 
on the transport. For large $\xi$'s, 
significant effects arise at Coulomb strengths $J$ substantially larger 
than $\Gamma$, while for $\xi \sim \beta$ values $J \sim \Gamma$ suffice. 

As mentioned from the very beginning, the main interest of the present work was not 
to analyze the physical properties of a simple model (i.~e., IRLM), 
to which all our numerical results 
refer. Rather, we attempted to extract from them information, which is useful   
from the perspective of real systems of interest for molecular transport.
Nevertheless, in spite of its simplicity, we emphasize that the IRLM incorporates 
a significant physical aspect, namely the Coulomb repulsion at the contacts.


Concerning the physical relevance of the employed IRLM, we also 
want to draw the attention on the following fact.
To give support to the robustness of their theoretical results, 
many investigators mentioned that reasonable changes of the molecule-electrode 
distance $d$ do not notably alter the calculated transport properties. 
Such a behavior can indeed be an indication for robust results, but it could 
also have a different origin. Within the IRLM, an increasing $d$ makes
both parameters $\beta$ and $J$ to decrease. The decrease of $\beta$ 
always causes a conductance reduction. As already noted and visible by inspecting 
the renormalized
curves of Fig.~\ref{fig:g_asym}, depending on the range,
a decrease in $J$ could either suppress or enhance the conductance. Noteworthy, 
this effect brought about by a modification in $J$ can be observed on the 
\emph{renormalized} curves. For realistic values of $J$($\sim 1$), the 
\emph{unrenormalized} 
conductance increases with decreasing $J$. So, rather than indicating
robust results, the aforementioned behavior could be related to the fact that 
the size of the extended molecule employed in the calculations is too
small, and for that size there is a compensation of 
two opposite effects induced by the $\beta$- and $J$-variations.

Of course, we do not expect that in all cases the largest extended molecule  
that can be tackled is sufficiently large to justify neglecting the influence
of electron correlations on the self-energies at the contacts. In Sect.~\ref{sec-renorm}, 
we considered a renormalization scheme appropriate for the IRLM, for which 
the exact conductance on resonance is known. To avoid misunderstandings,
we emphysize that this scheme represents an \emph{extra} renormalization,
which solely accounts for the fact that the parentheses of Eqs.~(\ref{eq-dyson-corr}) 
and (\ref{eq-dyson-work}) do not exactly vanish. 
The \emph{important} renormalization (the exact intramolecular correlations of the 
isolated extended molecule) is already incorporrated in
$\boldsymbol{G}_{i}^{r}$.
Simple schemes similar to our, which underline the renormalization considered 
in the Landau Fermi liquid theory, 
work pretty well even in Kondo context (see, e.~g., 
the supplementary material of Ref.~\onlinecite{tosatti:09}).
Certainly, the renormalization scheme of Sect.~\ref{sec-renorm}
cannot be applied in an identical form in general.  
However, we think that the possibility 
to renormalize the Green's function $\boldsymbol{G^{r}}(\varepsilon)$ 
of the embedded molecule by an $\varepsilon$-independent factor, which furthermore 
turned out to be (practically) independent on the model parameters, is interesting 
and deserves further attention. This (almost) constant factor can no more be determined from 
the unitary limit, as done in Sect.~\ref{sec-renorm}, but one may think to determine it 
by using general properties of the Green's functions (e.~g., sum rules or asymptotic behavior
\cite{mahan,tosatti:09}). 
The speculation on the existence of a (nearly) constant renormalization 
factor is also appealing from a different perspective. In the cases where the 
renormalization effect can be accounted for by an almost constant multiplicative factor of 
the Green's function, the renormalized and unrenormalized curves for current or linear 
conductance roughly differ by a nearly constant scaling factor 
[cf.~Eqs.~(\ref{eq-I}) and (\ref{eq-conductance})]. It is tempting
to relate this to the fact that, in spite of substantial 
quantitative discrepancies, this is just the basic difference between the 
calculated and measure curves observed in some cases.
\section{Conclusion}
\label{sec-conclusion}
In the present paper, we have presented a theoretical approach to correlated molecular 
transport relying upon a clear physical concept, namely that of a sufficiently large extended molecule.
The analysis of the present numerical results obtained for a simple but non-trivial correlated 
case has indicated that this approach can be legitimately applied to the molecular
transport in real systems. Computing the Green's functions 
within accurate quantum chemical calculations appears to be a feasible task for the isolated medium-size 
molecules needed. This is particularly true for narrower-band (organic) electrodes, but wider-band 
(metallic) electrodes can still be tackled especially in the most frequent situations of 
non-resonant tunneling. To remedy the stronger size-dependence for wider-band electrodes 
closer to resonance, 
we have suggested that simple renormalization procedures inspired by the Landau Fermi liquid theory 
can work, 
because it is unlikely that electron correlations in the systems of interest for molecular transport 
could invalidate the latter. 
Although our main interest was not the IRLM, we still note that the 
counter-intuitive theoretical prediction, 
that the contact electron-electron interaction could enhance the conductance, is 
interesting and can make the object of a parallel experimental investigation. 
\section*{Acknowledgments}
I.~B.~thanks Jochen Schirmer for valuable discussions and Uri Peskin for 
useful remarks. 
The financial support for this work 
provided by the Deu\-tsche For\-schungs\-ge\-mein\-schaft (DFG) is gratefully 
acknowledged.
%
%

\begin{thebibliography}{10}
\bibitem{Lambda-at-Fermi-energy}
Interestingly, at the Fermi energy, the correction factor
  ${\boldsymbol{\Lambda}}(\varepsilon_F) =\boldsymbol{1}$, and Eq.~(18) for the
  zero-bias conductance of Ref.~\onlinecite{Ferretti:05} becomes identical to
  the formula used for all the numerical results presented below. So, that
  equation is justified provided that the central region (extended molecule) is
  large enough.
\bibitem{N=1=3}
Strictly, we would have to take $N=1$ but then we would miss the Coulomb
  contact interactions.
\end{thebibliography}
%
%
%

%
\end{document}